\documentstyle[12pt]{article}
\renewcommand{\theequation}{\arabic{section}.\arabic{equation}}
\newcommand{\bm}[1]{\mbox{\boldmath $#1$}}
\newcommand{\be}{\begin{equation}}
\newcommand{\ee}{\end{equation}}
\newcommand{\ba}{\begin{eqnarray}}
\newcommand{\ea}{\end{eqnarray}}
\newcommand{\lb}{\label} 

\begin{document}
\begin{center}
\Large{ 
EHLERS--HARRISON TRANSFORMATIONS AND BLACK HOLES 
IN DILATON--AXION GRAVITY WITH MULTIPLE VECTOR FIELDS
%N=4 SUPERGRAVITY
}
\end{center}
\begin{center}
{\bf D. V. Gal'tsov\footnote{Permanent 
address: Department of Theoretical Physics, Physics Faculty, 
Moscow State University, 119899,  Moscow, Russia, 
email: galtsov@grg.phys.msu.su}
and  P. S. Letelier\footnote{e-mail: letelier@ime.unicamp.br}}
\\
Departamento de Matem\'{a}tica Aplicada -- IMECC\\
Universidade Estadual de Campinas\\                   
13081 Campinas, S.P., Brazil\\
\end{center}

\centerline{\bf Abstract}

\begin{quotation}
\vskip .5 cm
\noindent 
Dilaton--axion gravity with $p\;U(1)$ vector fields is studied
on space--times admitting a timelike Killing vector field.
Three--dimensional $\sigma$--model is derived in terms
of K\"ahler geometry, and  holomorphic representation of the 
$SO(2,2+p)$ global symmetry is constructed. A general static black hole
solution depending on $2p+5$ parameters is obtained via $SO(2,2+p)$
covariantization of the Schwarzschild solution. The metric in the curvature
coordinates looks as the variable mass Reissner--Nordstr\"om one 
and generically possesses two horizons. The inner horizon is pushed to 
the singularity if electric and magnetic $SO(p)$ charge vectors 
are parallel. For non--parallel charges the inner horizon has a finite
area except for an extremal limit when this property is preserved
only for orthogonal charges. Extremal dyon configurations with orthogonal 
charges have finite horizon radii continuously varying  from
zero to the ADM mass. New general solution is endowed with a NUT parameter,
asymptotic values of dilaton and axion, and a gauge parameter which can
be used to ascribe any given value to one of scalar charges.
\end{quotation}
\newpage
\baselineskip18pt

\section{Introduction}
\renewcommand{\theequation}{1.\arabic{equation}}
 
Recently there has been a substantial progress in understanding the 
statistical origin of the black hole entropy in the string 
theory (for a 
review and references see \cite{gho}). In achieving this goal it 
was important to realize that black holes in theories with dilaton
and multiple vector fields may have BPS saturated states with the 
finite area of the horizon. For toroidally compactified heterotic string
the effective four--dimensional theory is $N=4$ supergravity coupled 
to  Abelian vector multiplets. The above  feature is
manifest already at the level of the pure $N=4$ supergravity. 
Bosonic sector of this theory consists of the gravity coupled system 
of six $U(1)$ vector fields interacting with dilaton and axion. 
It is well known that, with only one $U(1)$ component
excited, the dilaton causes the horizon of the extremal black hole
to shrink to the singularity \cite{gi,ghs,cg}. With more than one $U(1)$ 
charges present, some BPS configurations may  possess 
the finite horizon area  like the Reissner--Nordstr\"om
solution. This was demonstrated by Kallosh {\it et al.} \cite{rk} 
within the two--vectors model without axion. Extremal
black holes with finite horizon area have $N=1$ residual
supersymmetry \cite{rk,bko} contrary to an extremal ``dilatonic''
black hole with vanishing horizon area, which exhibits two supersymmetries
unbroken \cite{rk4}. Similar conclusions were made in the heterotic
string effective theory with additional vector multiplets present \cite{cv}.

In view of the above it is important to investigate the space of classical 
solutions to pure and matter coupled $N=4,\, D=4$ supergravity 
in more detail. Although many particular black hole solutions 
were obtained earlier using various solution 
generating techniques, or solving Bogomol'nyi type equations in the BPS 
limit \cite{tod}, some recent study \cite{bko} revealed that this 
theory still requires further analysis. Of particular importance are 
stationary solutions which can be explored using powerful integration 
methods developed earlier in the vacuum General 
Relativity and Einstein--Maxwell theory \cite{ex}. Their generalization 
to dilaton--axion gravity interacting with one vector field
was  elaborated recently \cite{gk,diak}. The approach consist
in derivation of the three--dimensional $\sigma$--model, revealing the 
isometries of the target manifold (potential space), and utilizing its 
K\"ahler structure to achieve a more concise formulation. This procedure
has direct similarity with the Ernst approach to General Relativity 
\cite{er} based on the introduction of complex potentials parameterizing
K\"ahler target manifolds $SL(2,R)/SO(2)$ (vacuum) and
$SU(2,1)/(SU(2)\times U(1))$ (electrovacuum) \cite{mg}. It brings  
considerable simplifications into solution generating techniques 
both at the level of the finite--dimensional symmetry group, and  the 
infinite--dimensional Geroch--type
symmetries if further assumption of the second spacetime symmetry
(axial) is made \cite{bak,bdg}.  
 
K\"ahler parameterization of the target manifold of the stationary
dilaton--axion gravity with multiple vector fields was given
in \cite{ad}. The purpose of the present paper is to develop the 
corresponding solution generating technique and construct the most 
general non--rotating black hole solution to $N=4$ theory in a form 
manifestly covariant under the three--dimensional $U$--duality group.
We classify  isometries of the target manifold in  usual terms of 
General Relativity (Ehlers--Harrison transformations, electric--magnetic 
duality, gauge, scale transformations and $SO(p)$ rotations), and give 
explicitly the  corresponding holomorphic maps ($p$ being the 
number of $U(1)$ vector fields). 
Choosing particular combinations  which preserve asymptotic
conditions, we apply them to the Schwarzschild
solution to covariantize it with respect to the $U$--duality group
$SO(2,2+p)$. Thus new solution depending on $2p+5$ real
parameters is generated. 

Our solution includes many particular previously known black hole
configurations, and
allows one to determine their position in the general parameter space.
The metric depends on five parameters (including NUT), while the
material fields  are determined by two charge vectors,
asymptotic values of dilaton and axion,  and  a gauge
parameter which can be used to generate solutions with a
prescribed value of one of scalar charges. The metric is presented
in the curvature coordinates facilitating its physical interpretation and  
analysis of the BPS limit. It is shown that {\it generic non--extremal} 
black hole possesses two non--singular horizons and has a modified 
Reissner--Nordstr\"om form unless charges are parallel. 
If they are parallel, the internal horizon
shrinks to the singularity and the metric reduces to that of
the ``dilaton'' black hole \cite{gi,ghs}. In the case of orthogonal
charges of equal norm the metric is pure Reissner--Nordstr\"om
(thus we prove that this configuration found previously in \cite{rk} 
is unique). In the general case the metric deviates  from the usual 
Reissner--Nordstr\"om one by a ``variable mass'' term, asymptotically
equal to ADM mass and depending on one additional parameter. 
In the BPS limit the (NUT--less) solution with non--orthogonal charges 
has a ``dilatonic'' form (event horizon shrinking to the singularity).
When charges are orthogonal but have non--equal norms, the radius
of the event horizon in curvature coordinates varies between
zero value (for the non--dyon case) and the ADM mass (for the 
case of equal norms).

The plan of the paper is as follows. In the sec.~2 the 
four--dimensional theory is reviewed and 
the symmetries of the equations of motions are briefly discussed.
Then we perform $1+3$ decomposition of the metric and derive
a $\sigma$--model representation in three dimensions (sec.~3). 
Infinitesimal isometries of the target manifold are presented in 
the sec.~4. Complex parameterization of the target manifold 
is introduced in the sec.~5 where the holomorphic maps corresponding
to the target space isometries are constructed. The sec.~6 is devoted
to the covariantization of the Schwarzschild solution with respect
to $U$--duality of the three--dimensional theory.  In the sec.~7
we discuss the choice of coordinates, analyze various
particular black hole configurations and study the BPS limit. We 
conclude with brief remarks clarifying the relationship to  
previously done works.      

\section{ Preliminaries}
\renewcommand{\theequation}{2.\arabic{equation}}
\setcounter{equation}{0}
We consider the gravity coupled system on $p$ Abelian vector fields
$A_\mu^a,\; a=1,...,p$, one scalar $\phi$ (dilaton), and one
pseudoscalar $\kappa$ (axion) described by the action
\ba
&&S=\frac{1}{16\pi}\int \left\{-R+2\partial_\mu\phi\partial^\mu\phi +
\frac{1}{2} {\rm e}^{4\phi} 
{\partial_\mu}\kappa\partial^\mu\kappa
-{\rm e}^{-2\phi}F^a_{\mu\nu}F^{a\,\mu\nu}\right. \nonumber\\
&&\hspace{3cm}\left.-\kappa F^a_{\mu\nu}{\tilde 
F}^{a\,\mu\nu}\right\}\sqrt{-g}d^4x, \lb{s1}
\ea
where ${\tilde F}^{a\,\mu\nu}=\frac{1}{2}E^{\mu\nu\lambda\tau}
F^a_{\lambda\tau},\; F^a=dA^a,\; R$ is the (four--dimensional)
scalar curvature corresponding to  metric $g_{\mu\nu}$, sum over
repeated $a$ is understood. For $p=6$
this is the bosonic part of the $N=4, D=4$ supergravity.

Equations for vector fields following  from this action   have  a form
of modified Maxwell equations
\begin{equation} \lb{m1}
\nabla_\nu ({\rm e}^{-2\phi}F^{a\,\mu\nu}+\kappa{\tilde F}^{a\,\mu\nu})=0,
\end{equation}
together with the Bianchi identities
\begin{equation} \lb{m2}
\nabla_\nu{\tilde F}^{a\,\mu\nu}=0.
\end{equation}

The dilaton satisfies the curved space D'Alembert
equation with  axion and  vector sources:
\begin{equation}
\nabla_\mu\nabla^\mu\phi =\frac{1}{2}{\rm e}^{-2\phi}F^a_{\mu\nu}F^{a\,\mu\nu}
+\frac{1}{2}{\rm e}^{4\phi}(\nabla\kappa )^2,
\end{equation}
while the axion is generated by the pseudoscalar invariant
of vector fields
\begin{equation}
\nabla^\mu ({\rm e}^{4\phi}\nabla_\mu\kappa )=-
F^a_{\mu\nu}{\tilde F}^{a\,\mu\nu}.
\end{equation}
The system is closed by the Einstein equations with vector and
scalar sources
\begin{equation}
R_{\mu\nu}=2\nabla_\mu\phi\nabla_\nu\phi+
\frac{1}{2}{\rm e}^{4\phi}\nabla_\mu\kappa \nabla_\nu\kappa+
{\rm e}^{-2\phi}(2F^a_{\mu\lambda}{F^{a\,\lambda}}_\nu +
\frac{1}{2}F^a_{\lambda\tau}F^{a\,\lambda\tau}g_{\mu\nu}),
\end{equation}
where $(\nabla\kappa )^2\equiv g^{\mu\nu}
(\nabla_\mu\kappa)(\nabla_\nu\kappa)$, and $\nabla_\mu$
is the covariant derivative
with respect to the 4--dimensional metric $g_{\mu\nu}$.

The action (2.1) is invariant under the global $SO(p)$ rotations
of vector fields,
\begin{equation}
A^a_\mu \rightarrow {\Omega^a}_b A^b_\mu,\quad
{\Omega^a}_b {\Omega^b}_c =\delta^a_c.
\end{equation}
Another symmetry of the equations of motion, known as $S$--duality
\cite{sd}, can be concisely expressed using the complex axidilaton
variable
\be
z=\kappa+i{\rm e}^{2\phi}.
\ee
In terms of $z$ the action (2.1)  can be rewritten as
\begin{equation} \lb{s2}
S=-\frac{1}{16\pi}\int \left\{R+2 \nabla z \nabla{\bar z}(z-{\bar z})^{-2} -
2{\rm Re}\left(iz{\cal F}_{\mu\nu}^a{\cal F}^{a\, \mu\nu}\right)\right\}
\sqrt{-g}d^4x,
\end{equation}
where ${\cal F}^a=(F^a+i{\tilde F}^a)/2,\;
{\tilde F}^{a\, \mu\nu}=\frac{1}{2}{\rm e}^{\mu\nu\lambda\tau}
F^a_{\lambda\tau}$
(bar denotes complex conjugation).
Then the (four--dimensional) $S$--duality transformations read
\begin{equation}
z \rightarrow \frac{\alpha z+\beta}{\gamma z+\delta},\quad
{\cal F}_{\mu\nu}^a\rightarrow
(\gamma {\bar z}+\delta){\cal F}^{a\, \mu\nu},\quad
\alpha \delta -\beta \gamma =1.
\end{equation}
They leave invariant the kinetic term of the $z$--field in the action 
(2.9) as well as the full set of equations of motion 
(but not the action).

It is worth comparing the present theory with the Einstein--Maxwell
one (bosonic sector of $N=2$ supergravity):
\be
S=-\frac{1}{16\pi}\int \left\{R +
2\left({\cal F}_{\mu\nu}^a{\bar{\cal F}}^{a\, \mu\nu}\right)\right\}
\sqrt{-g}d^4x.
\ee
In this case both the Maxwell equations and Bianchi identities can be 
combined
into one complex equation
\begin{equation} \lb{m3}
\nabla_\mu{\cal F}^{a\,\mu\nu}=0,
\end{equation}
while the Einstein equations read
\be
R_{\mu\nu}=4 {\cal F}^a_{(\mu\lambda}{{\bar{\cal F}}^{a\,\lambda}}_{\nu)}.
\ee
Electric--magnetic duality now is an Abelian symmetry of the full action:
\be
{\cal F}^a_{\mu\lambda} \rightarrow {\rm e}^{i\alpha}
{\cal F}^a_{\mu\lambda},
\ee
where $\alpha$ is a constant parameter. Notice that 
the action of the dilaton--axion theory (2.9), 
which  has two additional field variables with respect
to (2.11), possesses at the same time larger symmetries 
(three--parametric $SL(2,R)$ instead of one--parametric $U(1)$). 
Due to this enhancement of four--dimensional symmetries, further 
three--dimensional reduction of the dilaton--axion gravity 
is describable as a non--linear $\sigma$--model on a symmetric 
space similarly to the Einstein--Maxwell theory.

\section{3+1 decomposition}
\renewcommand{\theequation}{3.\arabic{equation}}
\setcounter{equation}{0}
Let us consider the system of equations
(2.2-6) on a class of metrics depending effectively  on three
coordinates. In coordinate--independent way this can be expressed
as a restriction on spacetime to admit a Killing vector field.
It turns out that, if the Killing field is non--null everywhere,
the full system of equations can be presented in the form
of a gravity coupled non--linear sigma--model
on the symmetric space $SO(2, 2+p)/(SO(2)\times SO(2+p))$ 
or one of its non--compact form. Aiming to investigate black holes,
we assume here that the Killing 
vector is timelike in an essential region of spacetime.
Then ten components of $g_{\mu\nu}$ can be expressed using the standard
Kaluza--Klein ansatz through six components of the three--space
metric $h_{ij}$,  one (co)vector $\omega_i,\; (i, j=1, 2, 3)$,  
and one scalar $f$, depending on the space coordinates $x^i$:
\begin{equation}
ds^2=g_{\mu\nu}dx^\mu dx^\nu=f(dt-\omega_idx^i)^2-\frac{1}{f}h_{ij}dx^idx^j.
\end{equation}

To decompose vector fields, one has first to introduce the set of electric
potentials $A^a_0=v^a/\sqrt{2}$
(the coefficient helps to avoid undesired numerical factors in the
three--dimensional formulation).
The spatial parts of  four--potentials can be traded for another set
of scalars, magnetic potentials $u^a$, using the equations (2.2) for $\nu=i$.
Since  spatial   part of the tensor at the left hand side
is divergenceless, one  can write
\begin{equation}
{\rm e}^{-2\phi}F^{a\,ij}+\kappa {\tilde F}^{a\,ij}
=\frac{f}{\sqrt{2h}}\epsilon^{ijk}\partial_ku^a.
\end{equation}
This relation can be put into an alternative form
\be
{\tilde F}^a_{0i}=\frac{{\rm e}^{2\phi}}{\sqrt{2}}w^a_k,\quad
w^a_k =v(\partial_ku^a-\kappa\partial_kv^a),
\ee
and $w^a_k$ can be considered as covariant components of the 
three--dimensional vector ${\bf w^a}$.

The remaining Maxwell equations and  Bianchi identities now lead
to the following set of three--dimensional equations for $v^a,\, u^a$:
\begin{equation}
\bm\nabla\left(f^{-1}{\rm e}^{-2\phi} \bm\nabla v^a\right)
+f^{-2}\bm\tau \bm\nabla u^a-
\bm\nabla \left(f^{-1}\kappa
{\rm e}^{2\phi}{\bf w}^a\right)=0,
\end{equation}
\begin{equation}
\bm\nabla\left(f^{-1}{\rm e}^{2\phi}
{\bf w}^a\right)-f^{-2}\bm\tau\bm\nabla v^a=0,
\end{equation}
where  three--dimensional vector $\bm\tau$
dual to  two--form $d\omega$ is introduced
\begin{equation}
\tau^i=-f^2\frac{\epsilon^{ijk}}{\sqrt{h}}\partial_j\omega_k.
\end{equation}
Here (and in what follows) Latin indices are raised and lowered 
using the three--metric $h_{ij}$ and its inverse
$h^{ij}$, and $\bm\nabla$ denotes a three--space covariant derivative.

To clarify the role of $\bm\tau$, the mixed components
of the Ricci tensor should be invoked \cite{iw}:
\begin{equation}
R_0^i\equiv -\frac{f}{2\sqrt{h}}\epsilon^{ijk}\partial_k{\tau_j}.
\end{equation}
Upon substitution in  the source of  the corresponding components of the Einstein
equations, one finds
\be
 \epsilon^{ijk}\partial_k{\tau_j}=
2\epsilon^{ijk}\partial_jv^a\partial_ku^a.
\ee
This equation can be integrated by introducing the NUT (twist) potential 
$\chi$:
\begin{equation}
\tau_i=\partial_i\chi +v^a\partial_iu^a-u^a\partial_iv^a.
\end{equation}
Multiplying (3.9) by  $f^{-2}$, and taking into account (3.6),
one obtains the following second order equation for the NUT--potential:
\begin{equation}
f(\Delta\chi +v^a\Delta u^a-u^a\Delta v^a)=2(\bm\nabla \chi +
v^a\bm\nabla u^a
-u^a \bm\nabla v^a)\bm\nabla f,
\end{equation}
where
\begin{equation}
\Delta ={\bm\nabla}^2=\frac{1}{\sqrt{h}}\partial_i(\sqrt{h}
h^{ij}\partial_j)
\end{equation}
is a three--dimensional Laplacian.
 From  $R_{00}$ component of the Einstein equations one obtains an 
equation for the metric function $f=g_{00}$ in a similar form
\begin{equation}
f\Delta f-(\bm\nabla f)^2 +\bm\tau^2=
f\left({\rm e}^{2\phi}({\bf w}^a)^2+{\rm e}^{-2\phi}
(\bm\nabla v^a)^2\right).
\end{equation}
Finally, the dilaton and axion equations (2.4-5) reduce to
\begin{equation}
\Delta\phi =\frac{1}{2f}
\left({\rm e}^{-2\phi}(\bm\nabla v^a)^2-
{\rm e}^{2\phi}({\bf w}^a)^2\right)
+\frac{{\rm e}^{4\phi}}{2}(\bm\nabla \kappa)^2 ,
\end{equation}
\begin{equation}
\bm\nabla({\rm e}^{4\phi}\bm\nabla \kappa )=
\frac{2}{f}{\rm e}^{2\phi}{\bf w}^a \bm\nabla v^a,
\end{equation}

Now one can  check that this system of equations
is derivable  from the following three--dimensional action
\[
 S_m=\int \Biggl\{\frac{1}{2f^2}\left((\bm\nabla f)^2+
(\bm\nabla \chi +v^a\bm\nabla u^a-u^a\bm\nabla v^a)^2\right)
2(\bm\nabla\phi )^2+
\]
\be
+\frac{1}{2}{\rm e}^{4\phi}(\bm\nabla\kappa )^2-
\frac{1}{f}\left({\rm e}^{2\phi}({\bf w}^a)^2
+{\rm e}^{-2\phi}(\bm\nabla v^a)^2\right)\Biggr\}\sqrt{h}d^3 x,
\ee
which describes a non--linear $\sigma$--model with $2p+4$ scalar fields
 $\varphi^A=(f,\;$ $\chi,\;$ $v^a,\;$ $u^a,\;$ $\kappa,\;$ $\phi),$ $A=1,..., 2p+4$.
The remaining (spatial) Einstein equations reduce to three--dimensional
Einstein equations for the metric $h_{ij}$:
\begin{equation}
{\cal R}_{ij}=\frac{1}{2f^2}(f_{,i}f_{,j}+\tau_i\tau_j)+2\phi_{,i}
\phi_{j}+\frac{1}{2}{\rm e}^{4\phi}\kappa_{,i}\kappa_{,j}
-\frac{1}{f}\left({\rm e}^{-2\phi}v^a_{,i}v^a_{,j}+
{\rm e}^{2\phi}w^a_iw^a_j\right),
\end{equation}
where  ${\cal R}_{ij}$ is  three--dimensional Ricci tensor. 
Comparing with (3.15), one can see that
the source term  is derivable  from the same
action (3.15) as the  energy--momentum tensor. Therefore, the full
system of equations follows  from the three--dimensional gravity coupled
$\sigma$--model action
\begin{equation}
S_{\sigma}=\int \left(-{\cal R}+{\cal G}_{AB}
\partial_i\varphi^A\partial_j\varphi^B h^{ij}\right)\sqrt{h}d^3x,
\end{equation}
where the target space metric can directly read off  from (3.15) as follows
 \[
{dl}^2  = \frac{1}{2f^2}\left\{{df}^2+
(d\chi +v^adu^a-u^adv^a)^2\right\}+2{d\phi }^2
\]
\be
+ \frac{1}{2}{\rm e}^{4\phi }{d\kappa }^2
-\frac{1}{f}\left\{{\rm e}^{2\phi }(du^a-\kappa dv^a)^2
+{\rm e}^{-2\phi }(dv^a)^2\right\}.
\ee

The target manifold is a $2p+4$ dimensional pseudoeuclidean space of the
signature $2p-4$ which encodes the hidden global symmetries of the stationary
dilaton--axion gravity with $p$ vector fields ($p=1$ theory was studied 
earlier in \cite{gk}).
In absence of scalar fields it reduces (up to generalization 
to arbitrary $p$) to the potential space of Neugebauer and Kramer  
for the stationary Einstein--Maxwell system  \cite{nk}. It is worth noting,
however, that the target space of the Einstein--Maxwell theory does not
constitute a subspace of the full target manifold (3.18). Contrary to this,
the {\it vacuum} Einstein target manifold, parameterized by $f,\,\chi$  
{\it is}  a subspace. This means that there is an intrinsic connection
between stationary solutions of the vacuum (but not electrovacuum)
General Relativity and solutions
to the present theory.
\section{U--duality}
\renewcommand{\theequation}{4.\arabic{equation}}
\setcounter{equation}{0}
Global symmetries of the three--dimensional theory are manifest as
isometries of the target space.
As far as we have an explicit expression for the target space metric,
we can explore its isometries by solving Killing equations
\begin{equation}
K_{A;B}+K_{B;A}=0,
\end{equation}
where covariant derivatives refer to the  metric (3.18). General
solution to (4.1) was given in \cite{ggk} for a particular case $p=1$.
Generalization to arbitrary $p$ is rather straightforward in the
gauge and $S$--duality sectors, but is non--trivial in the
Ehlers--Harrison sector. Here we give the full set of 
solutions to (4.1) for arbitrary $p$. 

\vskip2mm
\noindent
{\bf Gauge transformations}
\vskip2mm

\noindent
This sector includes $2p+1$ linear constant shifts of electric and magnetic
potentials and a NUT potential. No physical quantities are changed.
Since electric and magnetic variables are mixed with NUT in (3.9), the
electromagnetic gauge transformations include appropriate variations
of  $\chi$
\begin{equation}
v^a\rightarrow v^a+e^a, \quad \chi \rightarrow \chi-u^a e^a,
\end{equation}
\begin{equation}
u^a\rightarrow u^a+m^a, \quad \chi \rightarrow \chi+v^a m^a,
\end{equation}
\be
\chi\rightarrow \chi+g,
\ee
 where $e^a,\, m^a$ and $g$ are real parameters. The corresponding Killing
vectors are
\begin{equation}
K^e_a=\partial_{v^a}-u^a\partial_{\chi },\quad
K^m_a=\partial_{u^a}+v^a\partial_{\chi },\quad
K^g=\partial_{\chi }.
\end{equation}

\vskip2mm
\noindent
{\bf Scale and SO(p) rotations}
\vskip2mm

\noindent
An invariance of (3.18) under rescaling
\begin{equation}
f\rightarrow {\rm e}^{2s}f, \quad
\chi\rightarrow {\rm e}^{2s}\chi,\quad
v^a\rightarrow {\rm e}^{s} v^a , \quad
u^a\rightarrow {\rm e}^{s} u^a
\end{equation}
is obvious, with $\kappa $ and  $\phi$ unchanged. This transformation
changes the four--dimensional metric.  The corresponding Killing vector is
\begin{equation}
K^s=2f\partial_f+2\chi \partial_{\chi }+v^a\partial_{v^a}+u^a\partial_{u^a}.
\end{equation}
Rotations in the space of $U(1)$ potentials (2.7) induces the 
corresponding transformations of electric and magnetic potentials:
\be
v^a \rightarrow {\Omega^a}_b v^b,\quad
u^a \rightarrow {\Omega^a}_b u^b.
\ee
They are generated by
\be
K_{ab}=v^a\partial_{v^b}-v^b\partial_{v^a}+
u^a\partial_{u^b}-u^b\partial_{u^a}.
\ee

\vskip2mm
\noindent
{\bf $S$--duality}
\vskip2mm

\noindent
Four--dimensional $S$--duality (2.10) induces the corresponding 
transformations
of the target space variables. They consist of the axion shift
\begin{equation}
\kappa\rightarrow \kappa+d_2,\quad u^a \rightarrow u^a+d_2 v^a,
\end{equation}
proper electric--magnetic rotations
\begin{equation}
u^a \rightarrow u^a, \quad v^a\rightarrow v^a+u^a d_1,\quad
z^{-1}\rightarrow z^{-1}+d_1,
\end{equation}
and $SL(2,R)$ scale transformation
\begin{equation}
z\rightarrow {\rm e}^{-2d_3} z,\quad
u^a \rightarrow {\rm e}^{-d_3}  u^a, \quad v^a\rightarrow 
{\rm e}^{d_3} v^a.
\end{equation}
The corresponding Killing generators
\begin{eqnarray}
& K^{d_1}=\partial_{\kappa }+v^a\partial_{u^a}, \quad
 K^{d_2}=({\rm e}^{-4\phi }-{\kappa }^2)\partial_{\kappa }+
\kappa \partial_ \phi +u^a\partial_{v^a},&\nonumber\\
& K^{d_3}=\partial_\phi -2\kappa \partial_{\kappa }+v^a\partial_{v^a}
-u^a\partial_{u^a}&
\end{eqnarray}
satisfy $sl(2,R)$ commutation relations
\begin{equation}
\left[K^{d_3},K^{d_1}\right]=2K^{d_1},\quad
\left[K^{d_3},K^{d_2}\right]=-2K^{d_2},\quad
\left[K^{d_1},K^{d_2}\right]=K^{d_3}.
\end{equation}

\vskip2mm
\noindent
{\bf Ehlers--Harrison sector}
\vskip2mm

\noindent
Non--trivial part of the isometry group includes
$2p$  Harrison transformations whose action on real
target space variables is rather involved (for $p=1$ case see
\cite{gk}). Fortunately, more concise form exists in terms of
complex variables, which we discuss in the next section. Here we
consider only  infinitesimal transformations. Note, that they are
not direct generalizations of those for $p=1$ \cite{gk}. 
The first set (electric) is
\begin{eqnarray}
&K_a^{H_1} = 2v^af\partial_f+v^a\partial_\phi+2w^a\partial_{\kappa }+
\left(f{\rm e}^{2\phi }-v^2\right)\partial_{v^a} +
2v^b\left(v^a\partial_{v^b}+u^a\partial_{u^b}\right)&\nonumber \\
& +\left(\chi-u^bv^b+\kappa f{\rm e}^{2\phi }\right)\partial_{u^a}+
\left(v^a(\chi+v^bw^b) +w^a(f{\rm e}^{2\phi }-v^2)\right)\partial_{\chi },&
\end{eqnarray}
where $v^2=(v^a)^2$, and we do not distinguish between upper and lower
$SO(p)$ indices.
The main effect of these transformations is to generate electric
potentials when acting on electrically uncharged configurations.
Their magnetic counterpart reads
\begin{eqnarray}
&\hspace{0.5cm}K_a^{H_2} = 2u^af\partial_f+(\kappa v^a-w^a)\partial_\phi+
2\left(\kappa w^a + v^a{\rm e}^{-4\phi }\right)\partial_{\kappa }+&\nonumber \\
&2u^b\left(v^a\partial_{v^b}+u^a\partial_{u^b}\right)-\left(u^bv^b+\chi
 -\kappa f{\rm e}^{2\phi }\right)\partial_{v^a} +
\left(f{\rm e}^{-2\phi }-u^2+
{\kappa }^2f{\rm e}^{2\phi }\right)\partial_{u^a}\nonumber &\\
&\hspace{0.5cm}+\left(u^a\chi +v^a(u^bw^b-f{\rm e}^{-2\phi })+
 w^a(\kappa f{\rm e}^{2\phi }-v^bu^b)\right)\partial_{\chi },&
\end{eqnarray}
where $w^a=u^a-\kappa v^a,\,u^2=(u^a)^2$.

The last, Ehlers--type \cite{eh} generator, which is produced
through the commutation relations
\begin{equation}
\left[K_a^{H_1},K_a^{H_2}\right]=2\delta_{ab}K^{E},
\end{equation}
reads
\begin{eqnarray}
&K^{E} = 2f\chi \partial_f+w^bv^b\partial_\phi+
(w^2-v^2{\rm e}^{-4\phi })\partial_{\kappa } &\nonumber\\
&+\left(v^b(\chi+v^aw^a) +w^b(f{\rm e}^{2\phi }-v^2)\right)\partial_{v^b} 
&\nonumber \\
& + \left(u^b\chi +v^b(w^2+\kappa v^aw^a-f{\rm e}^{-2\phi })+
 w^b(\kappa f{\rm e}^{2\phi }-\kappa v^2-v^aw^a)\right)\partial_{u^b}
 &\nonumber \\
&+\left({\chi }^2-f^2+fv^2{\rm e}^{-2\phi } +
w^2(f{\rm e}^{2\phi }-v^2)+(v^aw^a)^2\right)\partial_{\chi },&
\end{eqnarray}
where $w^2=(w^a)^2$.
Altogether these Killing vectors form a closed
$(p+3)(p+4)/2$--dimensional algebra.  To identify it with
$SO(2,p+2)$ let us consider a $4+p$--dimensional space  endowed
with the following pseudoeuclidean metric:
\begin{equation}
d\sigma^2=G_{\mu\nu}d\xi^\mu d\xi^\nu=
-(d\xi^0)^2-(d\xi^\theta)^2+(d\xi^1)^2+...+(d\xi^{p+2})^2,
\end{equation}
and denote the corresponding $so(2,p+2)$ generators as $L_{\mu\nu}$,
where $\mu, \nu= 0, \theta, 1,...,p+ 2$ and $a,\,b=1,...,p$. 
Then the following correspondence
between  $L_{\mu\nu}$ and the above Killing vectors can be established:
\begin{eqnarray}
&& L_{ab}=K_{ab},\;\;
   L_{0a}=\frac{1}{2}(K_a^{H_2}-K_a^m),\;\;
   L_{0\,p+1}=\frac{1}{2}(K^{d_3}-K^s),\nonumber\\
&& L_{0\,p+2}=\frac{1}{2}(K^g-K^{d_1}-K^{d_2}-K^E),\;\;
   L_{a\,p+1}=\frac{1}{2}(K_a^{H_2}+K_a^m), \nonumber\\
&& L_{p+1\,p+2}=\frac{1}{2}(K^g-K^{d_1}+K^{d_2}+K^E),\;\;
     L_{a\,p+2}=\frac{1}{2}(K_a^{H_1}+K_a^e),\nonumber\\
&& L_{p+1\, \theta}=\frac{1}{2}(K^g+K^{d_1}+K^{d_2}-K^E),\;\;
   L_{\theta\,p+2}=-\frac{1}{2}(K^{d_3}+K^s),\nonumber\\
&& L_{0\theta}=\frac{1}{2}(K^g+K^{d_1}-K^{d_2}+K^E),\;\;
   L_{\theta a}=\frac{1}{2}(K_a^{H_1}-K_a^e),
\end{eqnarray}

The target space (3.18) can now be identified with the coset manifold
$SO(2,p+2)/(SO(2)\times SO(2,p))$.

\section{Holomorphic representation}
\renewcommand{\theequation}{5.\arabic{equation}}
\setcounter{equation}{0}
The coset manifold $SO(2,p+2)/(SO(2)\times SO(2,p))$ may be
parameterized by $p+2$ complex coordinates and endowed
with the K\"ahlerian metric \cite{ad}. These coordinates can be regarded
as generalization of Ernst potentials of the Einstein--Maxwell theory.
The set consists of the axidilaton $z\,$, $p$ complex
electromagnetic potentials  
\be
\Phi^a=u^a-zv^a,
\ee
and the gravitational complex potential
\be
E=if-\chi+v^a\Phi^a.
\ee
In absence of vector fields (5.2) reduces to the original
Ernst potential $\epsilon=f+i\chi$ multiplied by $i$. However, when
there are no scalar fields, $E$ does not reduce to the Ernst potential
of the Einstein--Maxwell theory. In particular, $E$--potential is
not symmetric under an interchange $v^a\leftrightarrow u^a$ contrary
to its Einstein--Maxwell counterpart. This reflects an intrinsic asymmetry
of the $N=4$ supergravity with respect to electric and magnetic sectors.

In terms of complex variables
the target space metric (3.18) can be rewritten as
\ba
&dl^2=\frac{1}{2V^2}\left|{\rm Im}z\,dE+2{\rm Im}\Phi^a d\Phi^a-
\left({\rm Im}\Phi^a\right)^2\frac{dz}{{\rm Im}z}\right|^2-
\frac{1}{V}\left|d\Phi^a  \right. &\nonumber\\
&\left.-{\rm Im}\Phi^a\frac{dz}{{\rm Im}z} \right|^2+
\frac{1}{2}\left|\frac{dz}{{\rm Im}z}\right|^2,&
\ea
where the quantity
\be
V={\rm Im}E{\rm Im}z + \left({\rm Im}\Phi^a\right)^2
\ee
is related to the K\"ahler potential $K$  as follows \cite{ad}
\be
V={\rm e}^{-K}.
\ee
The K\"ahlerian metric (5.3) is generated by $K$ via
\be
\quad dl^2=2\left(\partial_\alpha \partial_{\bar\beta}K\right)\,
dz^\alpha d{\bar z}^\beta,
\ee
where holomorphic variables $z^\alpha$ are enumerated as follows
\be
z^\alpha =(E, \Phi^a, z),\quad \alpha=0,1,...,p+1.
\ee

Now the isometries of the target space may be presented as holomorphic
maps $z^\alpha \rightarrow z^{'\alpha}(z^\beta)$ leaving $K$ invariant
up to an admissible transformation
\be
K\rightarrow K + h(z) +{\bar h}({\bar z}),
\ee
where $h$ is an arbitrary holomorphic function of the K\"ahler coordinates.
First we describe two discrete maps which simplify substantially
the  derivation of holomorphic counterparts to  
continuous isometry transformations. One of them consists in the interchange
of the Ernst potential $E$ and the axidilaton:
\be
E\rightarrow z,\quad z\rightarrow E, \quad \Phi^a\;\; {\rm unchanged}.
\ee
This transformation does not modify $V$ (5.4) and hence leave the target
space metric unchanged. In terms of the real variables it corresponds to 
\be
f\rightarrow\frac{f}{f{\rm e}^{2\phi} -v^2},\quad
{\rm e}^{2\phi}\rightarrow f-v^2{\rm e}^{2\phi} ,\quad
v^a\rightarrow\frac{v^a}{f{\rm e}^{2\phi} -v^2},\quad
w^av^a-\chi \leftrightarrow \kappa
\ee
where $v^2=(v^a)^2$, while $u$ remains unchanged.
The second discrete transformation is more
sophisticated. It corresponds to an interchange of the Ernst and axidilaton
variables accompanied by the coordinate--dependent rescaling:
\be
E \rightarrow h^{-1}\,z, \quad
z \rightarrow h^{-1}\,E, \quad
\Phi^a \rightarrow h^{-1}\,\Phi^a,
\ee
where
\be
h(z^\alpha)=Ez+\Phi^2,\quad \Phi^2\equiv (\Phi^a)^2
\ee
is the function entering into the corresponding (admissible) transformation
of the K\"ahler potential (5.8). This transformation is rather complicated
being expressed in terms of real variables, but K\"ahler representation
makes it easy to check. This map has another
interpretation in the case $p=1$ \cite{diak,ad}  
as inversion of the ``matrix Ernst potential''.

Now we are in a position to list holomorphic maps corresponding
to continuous isometries of the target space.
Two simplest maps, linear shifts of $E$ and $\Phi^a$ on real constants
$g$ and $m^a$, correspond to gravitational
\be
E'=E+g,\quad {\Phi '}^a=\Phi^a,\quad z'=z,
\end{equation}
and magnetic
\begin{equation}
E'=E,\quad {\Phi '}^a=\Phi^a+m^a,\quad z'=z
\end{equation}
gauge transformation (4.3-4).
An electric gauge transformation looks somewhat more complicated:
\be
z'=z, \quad {\Phi '}^a=\Phi^a+e^a z,\quad
E'=E-2 e^a \Phi^a- (e^a)^2 z,
\ee
while the scale transformation (4.6) is simply rewritten as
\begin{equation}
E'={\rm e}^{2s}E,\quad {\Phi '}^a={\rm e}^s\Phi^a,\quad z'=z.
\end{equation}

Now apply the discrete map (5.9) to the electric gauge transformation (4.2).
Decomposing real Killing vectors of the previous section into holomorphic
and antiholomorphic parts one obtains 
\be
K^e_a=z\partial_{\Phi^a}-2\Phi^a \partial_E,
\ee
so that after (5.9) we get
\be
K^{H_1}_a=2\Phi^a \partial_z-E\partial_{\Phi^a},
\ee
what can be identified with the holomorphic part of the electric
Harrison Killing vector (4.15). This means that the finite electric Harrison
holomorphic map may be constructed via an interchange of $E$ and
$z$ in (5.15):
\begin{equation}
E'=E,\quad \Phi^a=\Phi^a+\nu^a E_,\quad
z'=z-2\nu^a \Phi^a-\nu^2 E,
\end{equation}
where $\nu^a$ is a set of real parameters, and $\nu^2=(\nu^a)^2$.

Similarly, applying the discrete map (5.11) to the magnetic gauge (4.3),
which infinitesimally reads
\begin{equation}
K^m_a=\partial_{\Phi^a},
\end{equation}
one obtains
\begin{equation}
K^{H_2}_a=(Ez+\Phi^2)\partial_{\Phi^a}- 2\Phi^a \left(z\partial_z +
E\partial_E\right)-2\Phi^a\Phi^b\partial_{\Phi_c}.
\end{equation}                             
This can be identified with the holomorphic part of the magnetic Harrison
Killing vector (4.16), and thus applying (5.11) 
to the finite magnetic gauge transformation (5.14) one obtains the 
holomorphic map corresponding to finite magnetic Harrison transformations
\begin{equation}
E'=\frac{E}{\Lambda},\quad
{\Phi '}^a=\frac{\Phi^a+\mu^a}{\Lambda},\quad
z'=\frac{z}{\Lambda},\quad
\Lambda=1+2 \mu^a \Phi^a+\mu^2 (E z+\Phi^2),
\end{equation}
where $\mu^2=(\mu^a)^2$. In the same way acting by (5.11) on the 
gravitational gauge
\begin{equation}
K_g=-\partial_E,
\end{equation}
one obtains Ehlers holomorphic generator
\begin{equation}
K^{E}=E\left(E\partial_E+\Phi^a\partial_{\Phi^a}\right)-\Phi^2\partial_z.
\end{equation}
The corresponding finite form is
\begin{equation}
E'=\frac{E}{1+c E},\quad
{\Phi '}^a=\frac{\Phi^a}{1+c E},\quad
z'=z+\frac{c\Phi^2}{1+c E}.
\end{equation}

The remaining $S$--duality transformations can be obtained in a similar
way by applying (5.9) to the gravitational gauge (5.13), scale (5.16) 
and   Ehlers transformations. This results in
\begin{eqnarray}
&& i)\qquad E'=E,\quad {\Phi '}^a=\Phi^a,\quad z'=z+b,\nonumber\\
&& ii)\qquad E'=E+\frac{d_1 \Phi^2}{1+d_1 z},\quad
{\Phi '}=\frac{\Phi^a}{1+d_1 z},\quad
z'=\frac{z}{1+d_1 z},\\
&& iii)\qquad E'=E, \quad {\Phi '}={\rm e}^{d_3}\Phi^a,\quad
z'={\rm e}^{2d_3}z.\nonumber
\end{eqnarray}

Thus we have constructed   $U$--duality finite $SO(2,2+p)$ transformations
in terms of holomorphic maps acting on complex potentials which can be 
regarded as Ernst--type variables. In the case
$p=1$ our holomorphic maps coincide with given previously
in \cite{diak}, where they were found in a different way using a
symplectic representation of the $so(2,3)$ algebra.
In the next section we will consider their application to solution
generation purposes.

%%%%%%%%%%%%%%%%%%%%%%%%%%%%%%%%%%%%%%%%%%%%%%%%%%%%%%%%%%%%%%%%%%%%%%

\section{ $SO(2,2+p)$ covariantization of the Schwarzschild solution}
\renewcommand{\theequation}{6.\arabic{equation}}
\setcounter{equation}{0}

Target space of the vacuum Einstein gravity $SL(2,R)/SO(2)$ can be
parametrized by a single complex variable $E_0=if-\chi$ and form
a subspace of (5.3). Using holomorphic transformations of the previous
section one can  construct a fully  $SO(2,2+p)$ covariant
counterpart to any vacuum solution of the Einstein equations.
An  application of this procedure to the
Schwarzschild metric is likely to produce a general
static black hole solution to dilaton--axion gravity with multiple
vector fields. Brief results were reported recently \cite{gl}.
Here we give more detailed derivation and discuss
further properties of the solution obtained.

In terms of holomorphic potentials  the
Schwarzschild solution reads
\be
E_0=if_0,\quad f_0=1-\frac{2M_0}{r_0},\quad z_0=1, \quad\Phi_0^a=0,
\ee
and the three--metric (which remains non--affected
by transformations) has non--zero components 
$h_{rr}=1,\, h_{\theta\theta}=f_0 r_0^2,\,
h_{\varphi\varphi}=f_0 r_0^2\sin^2\theta$. Let us first formulate the
asymptotic conditions for holomorphic variables ensuring
asymptotically flat (Taub--NUT) behavior of the solution.
It is worth noting that the NUT parameter
enters as one of the $SO(2,2+p)$ charges, so it has to be included in
order to maintain the $SO(2,2+p)$ covariance. 

If asymptotic values of the dilaton and axion are allowed to take 
arbitrary values $\phi(\infty)=\phi_\infty,\;\kappa(\infty)=\kappa_\infty$, 
the charges should be defined as follows  in the limit
$r\rightarrow \infty$:
\be
E\sim i\left(1-\frac{2{\cal M}}{r}\right),\quad
z\sim z_\infty-\frac{2i{\cal D}}{r}e^{-2\phi_\infty},\quad
\Phi^a \sim -\frac{i\sqrt{2}{\cal Q}^a}{r}e^{-\phi_\infty},
\ee
where $z_\infty=\kappa_\infty +ie^{-2\phi_\infty}$ and
three complex charges are introduced ${\cal M}=M+iN$, (ADM mass and NUT),
${\cal D}=D+iA$, (dilaton and axion), ${\cal Q}^a=Q^a+iP^a$
(electric and magnetic). Holomorphic transformations listed
in the previous section can be combined into several sets preserving
the asymptotic conditions (6.2) \cite{gl}. It is convenient to perform
first the charging (Harrison) transformations and then Ehlers and one of 
$S$--duality rotations  imposing a condition $z_\infty=i$, and afterwards
restore an arbitrary value of $z_\infty$ via  (5.26). So,
as an initial step, we apply Harrison transformations (first magnetic and
then electric) combining them with suitable magnetic   and
electric  gauge and axidilaton   rescaling to preserve
conditions $E(\infty)=z(\infty)=i,\, \Phi(\infty)=0$.
This results in
\[ E=\frac{im{\tilde n}_f+2q(1-f_0)}{n m_f},\quad
z=\frac{im n_f-2q(1-f_0)}{n m_f}, \]
\be
\Phi^a= \frac{(1-f_0)\left(\mu^a n+\nu^a(2q-im)\right)}{nm_f},
\ee
where
\ba
& q=\nu^a\mu^a,\;\;n=1-\nu^2,\;\; m=1-\mu^2,  &\nonumber\\
& m_f=1-\mu^2f_0,\;\;n_f=1-\nu^2f_0,\;\;{\tilde m}_f=f_0-\mu^2,\;\;
{\tilde n}_f=f_0-\nu^2.&
\ea
 From here one can extract the following expressions for the metric function
$f$ and the NUT potential:
\be
f=\frac{mnf_0}{m_fn_f},\quad \chi=\frac{q(f_0^2-1)}{m_fn_f}.
\ee
The latter has a non--vanishing Coulomb term in the asymptotic
expansion, and hence, according to (6.2), the solutions is asymptotically
Taub--NUT with $N=2qM_0/mn$. (One can check that this is the only effect
produced by $\chi$, short range terms in $\chi$ are exactly compensated
by the contribution of vector fields in (3.9), so the metric does not
contain other terms than NUT due to $\omega_idx^i$.)

The drawback of this solution is that the NUT parameter is not free but
is determined by charge parameters, and it is non--zero unless the $SO(p)$
vectors $\mu^a,\,$ $\nu^a$ are orthogonal, or one of them is zero.
A NUT parameter of the solution, however, can be made arbitrary by applying
the Ehlers transformation (5.25), what can be easily seen considering its 
action on the asymptotics (6.2). Similar role in the $S$--duality subgroup is
played by the map (5.26 ii): it rotates dilaton and axion charges 
in the same way as Ehlers map rotates the ADM mass and NUT.  Combining
both these rotations with gauge and scale transformations to preserve
asymptotic conditions \cite{gl}, one obtains for the complex variable $E$
the following expression
\be
E=\frac{2q(1-f_0)(1+bc)-n(cm_f+b{\tilde m}_f) +im\left({\tilde n}_f-
bcn_f\right)}{2q(1-f_0)(c-b)+n(m_f-bc{\tilde m}_f)+im(bn_f+c{\tilde n}_f)},
\ee
where $c$ is the parameter of the Ehlers transformation, and $b=d_2$ is
that of (5.26 ii). The corresponding values of the ADM mass and NUT 
parameter depend only on $c$:
\[
M=\frac{M_0}{mn(1+c^2)}\left\{(1-\mu^2 \nu^2)
(1-c^2)+4cq\right\},
\]
\be
N=\frac{2M_0}{mn(1+c^2)}\left\{c(1-\mu^2 \nu^2)
-(1-c^2)q\right\}.
\ee
Choosing $c$ to be the root (regular at $q=0$) of the equation
\be
q(c^2-1)+c(1-\mu^2\nu^2)=0,
\ee
one selects strictly asymptotically flat solutions without NUT. In this case
\be
M=\frac{M_0(1+c^2)}{mn(1-c^2)}\left(1-\mu^2\nu^2\right),
\ee
where
\be
c=-\frac{1-\mu^2\nu^2}{2q}+
\left\{1+\left((1-\mu^2\nu^2)/2q\right)^2\right\}^{1/2}.
\ee
If $M_0\ge 0$, the ADM mass of the new solution is non--negative
if $\mu^2\le 1,\, \nu^2\le 1,\, c^2\le 1$ (the limiting values should
be considered more carefully in view of the singularity ). These restrictions
on the parameters will be assumed so forth.
Alternatively, by an appropriate choice of $c$ one can construct 
massless solutions, these are non--trivial is $N\neq 0$.

Finally, arbitrary asymptotic values of dilaton and axion  can be generated
via (5.25 i, iii), the metric being unaffected. The resulting complex 
axidilaton field reads
\be
z=\kappa_\infty-e^{-2\phi_\infty}\frac{2q(1-f_0)(1+bc)+n(c{\tilde m}_f+bm_f)-
im(n_f-bc{\tilde n}_f)}{2q(1-f_0)(c-b)+n(m_f-bc{\tilde m}_f)+
im(bn_f+c{\tilde n}_f)}.
\ee
According to asymptotics (6.2) one obtains the following values of dilaton and
axion charges in terms of the parameters introduced:
\[
D=\frac{M_0}{mn(1+b^2)}\left\{(\mu^2 -\nu^2)(1-b^2)-4bq\right\},
\]
\be
A=\frac{2M_0}{mn(1+b^2)}\left\{b(\nu^2-\mu^2)-(1-b^2)q\right\}.
\ee
Hence  $b$ plays for $D$ and $A$ the same role as
$c$ for ADM mass and NUT charge. In particular, any of  scalar
charges $D,\;A$ may be  set zero by an appropriate choice of $b$.

The remaining set of $U(1)$ complex potentials is given by
\be
\Phi^a=\frac{e^{-\phi_\infty}(1+c^2)^{1/2}(1+b^2)^{1/2}(1-f_0)
\left\{\mu^a n+\nu^a(2q-im)\right\}}{2q(1-f_0)(c-b)+n(m_f-bc{\tilde m}_f)+
im(bn_f+c{\tilde n}_f)},
\ee
while the corresponding electric and magnetic charges are
\[
Q^a=\frac{\sqrt{2}M_0\left\{(n\mu^a+2 q \nu^a)(b+c)+m\nu^a(1-bc)\right\}}
{mn(1+b^2)^{1/2}(1+c^2)^{1/2}},
\]
\be
P^a=\frac{\sqrt{2}M_0\left\{(n\mu^a+2 q \nu^a)(1-bc)-m\nu^a(b+c)\right\}}
{mn(1+b^2)^{1/2}(1+c^2)^{1/2}}.
\ee
Altogether the solution contain $2p+5$ arbitrary parameters $M_0,$ $\mu^a,$
$\nu^a,$ $b,$ $c,$ $\kappa_\infty,$ $ \phi_\infty$. However, the number
 of independent
physical charges together with asymptotic values of dilaton and axion
is $2p+4$. Indeed, both dilaton and axion charges (as can be directly checked
using (6.7, 12, 14)) are determined by the equations
\ba
D &=& \frac{M\left\{(P^a)^2-(Q^a)^2 \right\} -2NQ^aP^a}{2(M^2+N^2)},
\nonumber\\
A &=& \frac{N\left\{(Q^a)^2-(P^a)^2 \right\} -2MQ^aP^a}{2(M^2+N^2)},
\ea
through electric and magnetic charges, mass and NUT parameter,
or, in the complex form,
\be
{\cal D}=-\frac{({\cal Q}^a)^2}{2{\cal M}},
\ee
and thus are not independent parameters. Meanwhile the uniqueness theorem
holds \cite{bgm} for harmonic maps to symmetric spaces (which is the present
case too) saying that the solution is uniquely determined by the
Coulomb charges. Moreover, the metric depends in fact only
on three combinations of $\nu^a, \mu^a$, namely $\mu^2, \nu^2, q=\mu^a\nu^a$
(the only scalar invariants which can be built  from  
$SO(p)$--vectors). Also it can be shown that the metric is $b$--independent.
Extracting the function $f$ from (6.6, 11, 13),  one finds
\be
f=\frac{f_0 mn (1+c^2)}{m_f n_f+c^2{\tilde n}_f{\tilde m}_f+2cq
(1-f^2_0)},
\ee
that is the metric is determined by $\nu^2, \mu^2, q, c, M_0$ and a NUT
parameter entering through $\chi$. All this indicates, that $b$ is 
a gauge parameter. It is, however,
useful  to keep it arbitrary if one wishes to get solutions
with a prescribed value of one of scalar charges. In particular,
for
\be
b=\frac{2q}{\nu^2-\mu^2}+\frac{\mu^2-\nu^2}{|\mu^2-\nu^2|}
\left\{1+\left(4q^2/(\nu^2-\mu^2)\right)^2\right\}
\ee
(as in (6.10) we choose the root which remains regular for all values of
the parameters) the dilaton charge will be zero.

The real potentials
$v^a, u^a, \phi, \kappa$ take simpler form in a ``symmetric'' gauge
$b=c$. Then the scalar fields are
\be
e^{-2(\phi-\phi_\infty)}=
\frac{m \left\{m_f n_f+c^2{\tilde n}_f{\tilde m}_f+2cq(1-f^2_0)\right\}}
{n \left(m_f^2+c^2{\tilde m}_f^2\right)},
\ee
\be
\kappa-\kappa_\infty=
\frac{(f_0-1)\left\{c m(\mu^2-\nu^2) (1+f_0)+2q (m_f-c^2 {\tilde m}_f)\right\}}
{n\left(m_f^2+c^2{\tilde m}_f^2\right)},
\ee
while the $U(1)^p$ potentials read
\be
v^a=\frac{e^{\phi_\infty}(f_0-1)\left\{(c^2 {\tilde m}_f-m_f)\nu^a
-c(1+f_0)(\mu^a n+2q\nu^a)\right\}}
{m_f n_f+c^2{\tilde n}_f{\tilde m}_f+2cq
(1-f^2_0)},
\ee
\be
w^a=\frac{e^{-\phi_\infty}(f_0-1)\left\{(c^2 {\tilde m}_f-m_f)(\mu^a n+2q\nu^a)
 +c (1+f_0)\nu^a m^2\right\}}
{n\left(m_f^2+c^2{\tilde m}_f^2\right)},
\ee
where we have chosen  $w^a=u^a-\kappa v^a$  instead of $u^a$
which are pure magnetic asymptotically  when 
$\phi_\infty\neq 0,\, \kappa_\infty\neq 0$.

The solution (6.17, 19-22) constitutes the $SO(2,2+p)$ covariantization
of the Schwarzschild solution to which it reduces when all parameters
$\mu^2, \,$  $\nu^2, \,$ $ c, \,$  $\phi_\infty, \, \kappa_\infty$ are set zero.
It contains a maximal number of $U$--duality parameters compatible
with asymptotic conditions and up to $2p$ trivial constants which can be added
to $v^a,\,u^a$. It has arbitrary values of $2p$ vector charges, ADM mass,
NUT parameter, and the scalar charges which are determined in terms of
other charges via (6.15). This latter condition seems to be
necessary for the absence of naked singularities (although no general
proof has been given so far). Thus, by the uniqueness theorem,
which holds for harmonic map into symmetric spaces \cite{bgm}, it is likely
to be the most general black hole solution (possibly endowed with NUT)
to $N=4,\,D=4$ supergravity.

%%%%%%%77777777777777777777777777777777777777777777777

\section{Dilaton--Axion--Reissner--Nordstr\"om black holes}
\renewcommand{\theequation}{7.\arabic{equation}}
\setcounter{equation}{0}

Now let us discuss the choice of coordinates for the transformed metric.
One reasonable choice
is motivated by the Garfinkle--Horowitz--Strominger gauge for a (non--dyon)
dilatonic black hole \cite{ghs}, in which the curvature singularity sits
at some finite radius $r=r_-$:
\be
h_{\theta\theta}=f_0 r_0^2=fr(r-r_-).
\ee
Taking into account that the three--space metric $h_{ij}$ is not changed by
 transformations, one obtains for $r$ an equation
\be
r(r-r_-)=(r_0-r_0^+)(r_0-r_0^-),
\ee
where
\be
r_0^\pm=\frac{M_0}{mn(1+c^2)}\left\{2\alpha-\beta
\pm\left(\beta^2-4\alpha\gamma\right)^{1/2}\right\},
\ee with
\be
\alpha=c^2-2cq+\mu^2\nu^2,\;\;\beta=(\mu^2+\nu^2)(1+c^2),\;\;
\gamma=1+2cq+c^2\mu^2\nu^2.
\ee
Clearly
\be
r=r_0-r_0^-, \quad r_-=r_0^+-r_0^-,
\ee
and hence the four--dimensional metric reads
\be
ds^2= f(dt-2Nd\varphi)^2-f^{-1}dr^2
-r(r-r_-)\left(d\theta^2+\sin^2\theta d\varphi^2\right),
\ee
where
\be
f=1-\frac{2M}{r}+\frac{K}{r(r-r_-)},\quad -K=r_0^+(r_0^-+2M).
\ee
Here $M$ is a physical mass given by (6.7). The following relation
between the seed and physical masses is useful:
\be
M_0-M=(r_0^++r_0^-)/2,
\ee

Unless $K=0,\;g_{00}$ is not Schwarzschildean, and the metric (6.28) does not
coincide with that of \cite{ghs}. Hence
in the general case the gauge (6.23) has no advantage, and one can choose
more familiar curvature coordinates imposing a condition
\be
h_{\theta\theta}=f_0 r_0^2=f\rho^2.
\ee
Then the relation between the old and new coordinates is quadratic, and one
obtains
\be
r_0=\frac{1}{2}\left\{r_0^++r_0^-+\left(r_-^2+4\rho^2\right)^{1/2}\right\}.
\ee
In the curvature coordinates the transformed metric is rather similar to the
Reissner--Nordstr\"om one
\be
ds^2=\left(1-\frac{2Mg}{\rho}+\frac{C}{\rho^2}\right)
(dt-2Nd\varphi)^2-
\left(1-\frac{2Mg}{\rho}+\frac{C}{\rho^2}\right)^{-1}d\rho^2-
\rho^2 d\Omega^2,
\ee

with the same  mass $M$ and  ``charge'' parameter
\be
C=\frac{r_0^-}{2}(r_0^--2M_0)+\frac{r_0^+}{2}(r_0^+-2M_0).
\ee
It differs, however,  from the usual Reissner--Nordstr\"om--NUT metric by
a ``variable mass'' factor
\be
g=\left\{1+\left(r_-/(2\rho)\right)^2\right\}^{1/2},
\ee
which tends to unity when $\rho\rightarrow \infty$. If $r_-=0$, the solution
is pure Reissner--Nordstr\"om--NUT.  In this case, in view of (7.8), $K=C$,
and hence  (7.6) gives the same result. In what follows we will refer
to the metric (7.11) as Dilaton--Axion--Reissner--Nordstr\"om--NUT
(DARN--NUT).

Generally, the equation $g_{00}=0$ has two solutions, $\rho_{\pm}$. Taking
into account (7.9), one can realize that smaller root
corresponds to $r_0=0$, while bigger one to $r_0=2M_0$.
This means that
\be
\rho_{-}^2=r_0^-r_0^+,\quad
\rho_{+}^2=(r_0^--2M_0)(r_0^+-2M_0),
\ee
or, substituting (6.25), \be
\rho_{\pm}^2=\frac{4M_0^2}
{mn(1+c^2)}\pmatrix{c^2\mu^2\nu^2+1+2cq\cr
\mu^2\nu^2+c^2-2cq\cr}.
\ee
Let us consider now the regular DARN black hole,  assuming $c$ to obey
(6.10), so that $N=0$. Parameter $r_-$ entering the above formulas then
will be
\be
\frac{r_-}{2M}=\frac{\left\{ (1-c^2)^2(\mu^2-\nu^2)^2+
4c^2(1-\mu^2\nu^2)^2\right\}^{1/2}}{(1+c^2) (1-\mu^2\nu^2)},
\ee
while radii of the horizons are
\be
\rho_\pm=\frac{2Mm^{1/2}n^{1/2}(1-c^2)^{1/2}}{(1+c^2)(1-\mu^2\nu^2)}
\pmatrix{\left(1-c^2\mu^2\nu^2\right)^{1/2}\cr
\left(\mu^2\nu^2-c^2\right)^{1/2}\cr}.
\ee
The solution does not contain naked singularities if
$\mu^2\nu^2\le 1,\;c^2\le \mu^2\nu^2$ (what will be assumed below).

The inner horizon is pushed to the singularity when $c^2=\mu^2 \nu^2$.
In this case the solution is NUT--less if $q=c$. Both relations together
imply that in this case the $SO(p)$ vectors $\mu^a,\, \nu^a$  are parallel.
It can also be checked that $r_0^+=0$ and hence $K=0$, in which case we
recover the standard expression for the metric of a ``dilaton black hole''
\cite{ghs} with   singularity at $r=r_-$,
\be
r_-=\frac{2M(\mu^2 + \nu^2)}{1+\mu^2\nu^2}=
\frac{(Q^a)^2+ (P^a)^2}{M}.
\ee
The electric and magnetic charge vectors are also parallel and in the 
gauge $b=c$ read
\be
Q^a=\frac{\sqrt{2}\,M_0(1+\mu^2) \nu^a}{nm},\quad
P^a=\frac{\sqrt{2}\,M_0\mu^a}{m},\quad
M=\frac{M_0(1+\mu^2\nu^2)}{mn}.
\ee
The dilaton and axion charges are still given by general formulas
(6.15) (with $N=0$) which  now read
\[
D=\frac{(P^a)^2-(Q^a)^2}{2M}=
\frac{M\left\{\right(\mu^2-\nu^2)(1-\mu^2\nu^2)-4\mu^2\nu^2\}}
{(1+\mu^2\nu^2)^2},
\]
\be
A=-\frac{Q^aP^a}{M}=-\frac{2M(1+\mu^2)n\mu\nu}{(1+\mu^2\nu^2)^2},
\ee
where $\mu,\,\nu$ are norms of the vectors $\mu^a,\,\nu^a$. Note, that
in the case of a single ($p=1$) vector field $q=\mu\nu$ inevitably, and the
solution is NUT--less if $q=c$, what is just the present case. Therefore,
the generalization of the ``dilaton black hole'' solution to the case of
multiple vector fields is DARN solution with parallel
$SO(p)$ vectors $\mu^a,\, \nu^a$, (or, equivalently, parallel  $Q^a,\, P^a$).
 It can be presented in the curvature coordinates as (6.33),
where now
\be
C=Mr_-=(Q^a)^2+ (P^a)^2,
\ee
while the radius of the event horizon is
\be
\rho_+=2M\left(\frac{mn}{1+\mu^2\nu^2}\right)^{1/2}.
\ee

Now  let us clarify  conditions, under which DARN solution becomes
{\it pure} Reissner--Nordstr\"om. This happens when $r_-=0$.  From (7.16) it
is clear that one should have $\mu=\nu,\, c=0$ (thus implying
$q=0$ for vanishing NUT). Therefore the only occurrence of the
proper Reissner--Nordstr\"om solution in the dilaton--axion gravity with
multiple vector fields is the case of the  orthogonal
$SO(p)$ vectors $\mu^a,\, \nu^a$ of equal length.  In this case
\be
Q^a=\frac{\sqrt{2}M m\nu^a}{1-\mu^2\nu^2},\quad
P^a=\frac{\sqrt{2}M n\mu^a}{1-\mu^2\nu^2},
\ee
and the ``charge''parameter in (6.11) assumes its standard form
$C=(Q^a)^2+ (P^a)^2$.  From (6.19), (6.20) it is seen that in this case
\be\phi=\phi_\infty,\quad \kappa=\kappa_\infty
\ee
(``frozen moduli'' \cite{gkk}). The fact that equal electric and
magnetic charges belonging to different $U(1)$ sectors of
$N=4$ supergravity generate  the Reissner--Nordstr\"om metric
was discovered some time ago \cite{rk} within a truncated model
with no axion. Our results show that this
is the {\it  unique} configuration of the $N=4,\, D=4$ supergravity
(and its arbitrary-$p$ generalization) ensuring such a property.
In particular, if one takes  $c=b=0$ (in which case we go back
to the intermediate form  (6.3) of the solution up to modifications
due to an arbitrary asymptotic value $z_\infty$), and  $q=0$
to have  $N=0$, then
\be
r_-=\frac{2M\left|\mu^2-\nu^2\right|}{1-\mu^2\nu^2},
\ee and
\be
K=\left(\frac{2M}{1-\mu^2\nu^2}\right)^2
\pmatrix{m^2\nu^2,\;\;\mu^2>\nu^2\cr
n^2\mu^2,\;\;\nu^2>\mu^2\cr}
\ee
so the metric is neither Reissner--Nordstr\"om, nor dilatonic
if $\mu\ne\nu$ and $\mu,\nu\ne 0$. However, it {\it is} dilatonic
if either all electric, or all magnetic charges are zero (then
$K=0$).  From this formula one can see that there is another case
when $K=0$, namely, when either $\mu$ or $\nu$ is approaching unity. This
is the BPS limit, which is worth to be discussed separately.

Using formulas for physical charges in terms of the parameters
one can show that the following identity holds generally
\be
M^2+N^2+D^2+A^2-(Q^a)^2-(P^a)^2=M_0^2.
\ee
Thus the BPS limit of the solution corresponds to $M_0\rightarrow 0$.
 From (6.9) it is clear that in the NUT--less case one possibility
to achieve this limit (for non--zero $M$) is to take one  of two
quantities $\mu^2,\, \nu^2$ approaching unity and the other still keeping
some different value. Then $c\ne 1$, and $r_-\rightarrow 2M$. In the
limit one gets the following BPS saturation  conditions  
\be
(Q^a)^2+(P^a)^2=2M^2,\quad D^2+A^2=M^2.
\ee
(we choose the gauge $b=c$ in (6.12)).  From (7.17) it also follows that
both horizon radii shrinks to the singularity. This is the ``dilatonic''
BPS state which corresponds in $N=4$ supergravity ($p=6$) to $1/2$ unbroken 
supersymmetries \cite{bko}. Indeed, absolute values of
 two central charges in our variables read
\be
|z_{1,2}|^2=\frac{1}{2}\left\{(Q^a)^2+(P^a)^2
\pm\sqrt{(Q^a)^2(P^a)^2-(Q^a P^a)^2}\right\}.
\ee
For parallel charge vectors the square root vanishes, so both central charge
moduli are equal, and from the saturation condition $|z_{1,2}|=M$ with
account for (7.20) one obtains (7.28).

If {\it both} $\mu^2,\, \nu^2$ tends to unity simultaneously, then 
three cases should be distinguished:

i)  $\mu^a,\,\nu^a$ are parallel.

In this case $q=\mu\nu$, implying $c=\mu\nu$ to eliminate NUT, 
and we come back to the general dilatonic case discussed above.

ii) $\mu^a,\,\nu^a$ are neither parallel, nor orthogonal.

 From (6.10) one finds that then $c\rightarrow 1$ so that
\be
\frac{1-\mu^2\nu^2}{1-c^2}\rightarrow q,\quad q\ne 0,
\ee
and then $r_-\rightarrow 2M$, $\rho_\pm\rightarrow 0$, hence the
limit is also dilatonic.

iii) $\mu^a,\,\nu^a$ are  orthogonal.

In this case $q=0,\,c=0$, and both horizons merge at
\be
\rho_\pm=\rho_{ext}= \frac{2M\sqrt{mn}}{1-\mu^2\nu^2}.
\ee
The limit of the right hand side depends on the curve in the $\mu,\,\nu$ 
plane along which the point $\mu=1,\,\nu=1$ is reached. If one takes
the limit along the line $\mu=\nu$, one gets
\be
\rho_{ext}\rightarrow  M,
\ee
what corresponds to the extreme Reissner--Nordstr\"om black hole.
If one goes to the limiting point at some different angle, e.g.
\be
\mu=1-\epsilon,\quad \nu=1-k\epsilon,\quad \epsilon \rightarrow 0, 
\ee
one generally gets an extremal solution with the horizon radius
\be
\rho^2_{ext}=\frac{4kM^2}{(1+k)^2}.
\ee
In two limiting cases, $k=0,\, k=\infty$ (non--dyons)
the radius tends to zero,
hence solution is dilatonic. For $k=1$ it is the
extremal Reissner--Nordstr\"om, for all other values of $k$ this is an
extremal limit of the generic solution (7.11). Indeed, the parameter
$r_-$ determining the deviation  from the Reissner--Nordstr\"om geometry
now is given by
\be
r_-=\frac{2M|1-k|}{1+k},
\ee
the charge parameter is $C=(Q^a)^2+(P^a)^2$, while its ratio to
the square of mass is
\be
\frac{(Q^a)^2+(P^a)^2}{M^2}=\frac{2(1+k^2)}{(1+k)^2}.
\ee
The dilaton charge is equal to
\be
D=\frac{(P^a)^2-(Q^a)^2}{2M}=\frac{2M^2(k^2-1)}{(1+k)^2},
\ee
while the axion charge is zero, so the force balance condition
is satisfied
\be
M^2+D^2=(P^a)^2+(Q^a)^2.
\ee
Furthermore, the electric and magnetic charge vectors have the form
\be
Q^a=\frac{\sqrt{2}\,M\nu^a}{1+k},\quad
P^a=\frac{\sqrt{2}\,M\mu^a k}{1+k},
\ee
where now $\mu^a,\,\nu^a$ are unit vectors. Hence $k$ can be expressed
through the  norms of charge vectors
$Q=(Q^aQ^a)^{1/2},\,P=(P^aP^a)^{1/2}$ as
\be
k=\frac{P}{Q}.
\ee
 From (7.38) it also follows that
\be
P+Q=\sqrt{2}M,
\ee
and then, in view of (7.36-37),
\be
P-Q=\sqrt{2}D,
\ee
thus only one  BPS bound of $N=4$ supergravity is saturated, 
signalizing that the solution has $N=1$ residual supersymmetry \cite{rk,bko}.

Therefore, in the BPS limit the DARN solution with orthogonal
charge vectors has the horizon radius (7.33) taking all values in the interval
\be
0\le\rho_{ext}\le M.
\ee
Zero value is reached for purely electric $k=0$ or purely
magnetic $k=\infty$ configurations, while the upper limit corresponds
to a symmetric dyon $Q=P$.

To avoid confusion, note that this interpretation is based on our choice
of curvature coordinates (like in the standard  Reissner--Nordstr\"om
metric). In \cite{rk} it was claimed that {\it all} 
BPS dyonic black holes in purely
dilaton gravity with two vector fields, possessing electric and magnetic
charges in different $U(1)$ sectors, have the same radius of the horizon
equal to that of a  pure  Reissner--Nordstr\"om black hole. This may seem
to disagree with our conclusion since the particular model of \cite{rk}
lies within the scope of our consideration. There is no contradiction,
however, since in \cite{rk} a different coordinate system was used, namely
their radial variable (denoted by prime) is related to our $r$ (7.1)
as $r'=r-r_-/2.$ Taking into account (7.35), one can easily show that
in the BPS limit $r'=M$ independently on $k$. But the geometrical radius of
2-spheres in terms of primed variable is equal to
$\{r'^2-(r_-/2)^2\}^{1/2}$, whereas it is $\rho^2$ in our case.
Thus one has the same expression for the area of the horizon surface: 
\be
{\cal A}_H=4\pi \rho^2_{ext}=8\pi PQ.
\ee
Recall that this expression is valid only in the case of orthogonal charge
vectors. If desired, the solution (7.11) may be presented in the form
similar to  \cite{rk}. However, we prefer to use
curvature coordinates because of their more transparent
geometrical meaning.

\section{Conclusions}
We have given a concise representation of the three--dimensional $U$--duality
for the (generalized) bosonic sector of the stationary $N=4$ supergravity
in terms of K\"ahler parameterization of the target manifold
$SO(2,p+2)/(SO(2)\times SO(2,p))$, where $p$ is a number (chosen arbitrary 
for generality) of $U(1)$ vector fields. Global $SO(2,p+2)$ symmetry
of the three--dimensional theory was described in terms of holomorphic
maps exhibiting the target space isometries. They include Ehlers--Harrison
transformations, $S$--duality, as well as $SO(p)$ rotations, gauge and
scale transformations. Selecting transformations which preserve an asymptotic
flatness, one obtains a $(2p+4)$--parametric subset which is suitable
to perform a $SO(2,p+2)$ covariantization of any asymptotically flat
solution to the vacuum Einstein equations. Such covariantization
of  the Schwarzschild solution presents the most general static black hole 
in the dilaton--axion gravity with multiple vector fields.

Generically DARN black hole solution obtained possesses two horizons. 
The metric depends on four parameters (in the NUT--less case): 
the ADM mass, and three
$SO(p)$ invariants built out of two $SO(p)$ charge vectors. For
parallel charges  an internal horizon shrinks to the singularity
and the metric coincides with that of the standard dilaton black hole. 
For orthogonal charge vectors of equal length one has a pure 
Reissner--Nordstr\"om metric. All other charge configurations correspond to
generic DARN metric with two non--singular horizons and a space--like 
singularity. The NUT--generalization of the solution is also given.

Having  now  general non--extremal solution, one can  better
understand how generic are particular configurations found
earlier in the Bogomol'nyi--Gibbons--Hull limit \cite{rk,rk4,cg,bko}.
One finds that in the BPS limit  general DARN solution has a
``dilatonic'' behavior
(both horizons shrinking to the singularity), unless charge vectors
are orthogonal. Thus the generic BPS static black hole in the
$N=4,\, D=4$ supergravity has a vanishing area of the horizon (and N=2
residual SUSY). In the case of orthogonal charge vectors
two horizons merge at some (generically finite) radius depending
on the ratio of norms of the
electric and magnetic charge vectors. Non--dyon configurations
are still dilatonic, while dyons have finite radius of the horizon,
taking maximal value in the ``symmetric'' case.

Few remarks are in order concerning the relationship between our solution
generating technique and those, frequently used in the string theory,
which are usually based on the presentation of the  four--dimensional theory 
as dimensional reduction of some higher--dimensional one. First of all,
our direct construction of the sigma--model  
does not require the theory to be obtainable via dimensional reduction.
Also, the standard reduction procedure usually
gives rise to some  matrix representation in terms of real variables
which is manifestly invariant under action of the symmetry group, but physical
interpretation of transformations often remains obscure. 
The advantage of the present approach is that
it deals with complex potentials, thus reducing the total number
of variables by half. It also provides  from the very beginning for a clear 
physical identification of transformations involved. In addition, 
as we have shown, some discrete symmetries
exist which are manifest only in terms of K\"ahlerian variables.

\vskip3mm
{\large \bf Acknowledgments}\\
This work was supported in part by FAPESP and CNPq, Brazil.  D.G.  is
grateful to Departamento de Matematica Aplicada , IMECC,
Universidade Estadual de Campinas for hospitality during his visit.
The work of D.G. was also supported in part by the RFBR
Grant 96--02--18899.


\begin{thebibliography}{99}
\bibitem{gho}
G.T. Horowitz, {\it The Origin of Black Hole Entropy in String Theory},
UCSBTH--96--07, Gr-qc/9604951.

\bibitem{gi}
G.W. Gibbons, Nucl. Phys. {\bf B207}, 337 (1982);
G.W. Gibbons and K. Maeda, Nucl. Phys. {\bf B298}, 741 (1988).
\bibitem{ghs}
D. Garfinkle, G.T. Horowitz, and A. Strominger Phys. Rev. {\bf D43}, 3140
(1991); {\bf D45}, 3888 ({\bf E}) (1992).
\bibitem{cg}
G. Cl\'ement and D. Gal'tsov,
{\it Stationary BPS solutions to dilaton--axion gravity}, Preprint
GCR--96/07/02, DTP-MSU/96--11, Hep-th/9607043.

\bibitem{rk}
R. Kallosh, A. Linde, T. Ortin, A. Peet, and A. van Proyen,
Phys. Rev. {\bf D 46}, 5278 (1992).

\bibitem{bko}
E. Bergshoeff, R. Kallosh, and T. Ortin {\it Stationary
Axion/Dilaton Solutions and Supersymmetry}, Preprint UG--3/96,
SU--ITP--19, CERN--TH/96--106, Hep-th/9605059.
\bibitem{rk4}
R. Kallosh, D. Kastor, T. Ortin, and T. Torma,
Phys. Rev. {\bf D 50}, 6374 (1994).
\bibitem{cv}
M. Cveti{\v c} and D. Youm, {\it General Static Spherically Symmetric
 Black Holes of Heterotic String on a Six Torus},
IASSNS--HEP--95--107, Hep-th/9512127;
M. Cveti{\v c} and A. Tseytlin, Phys. Rev. {\bf D 53}, 5619 (1996);
M. Cveti{\v c} and A. Tseytlin, {\it Non--Extreme Black Holes  from
Non--Extreme Intersecting M--branes}, DAMTP/R/96/27, Imperial/TP/95--96/52, 
Hep-th/9606033;
M. Cveti{\v c} and C.M. Hull, {\it Black Holes and U--Duality},
DAMTP/R/96/31, QMW-96--12, Hep-th/9606193.
\bibitem{tod}
K.P. Tod, Phys. Lett. {\bf B 121}, 241 (1983); Class.~Quantum Grav.~{\bf 12}, 
1801 (1995).
\bibitem{ex}
D. Kramer, H. Stephani, M. MacCallum, and E. Herlt, {\it Exact
Solutions of the Einstein Field Equations\/}, CUP, 1980.
\bibitem{gk}
D.V. Gal'tsov and O.V. Kechkin,  Phys. Rev. {\bf D50},
7394 (1994); D.V. Gal'tsov, Phys. Rev. Lett. {\bf 74}, 2863 (1995);
D.V. Gal'tsov, {\em``Symmetries of Heterotic String Effective Theory
in Three and Two Dimensions''}, in
{\sl Heat Kernel Techniques and Quantum Gravity},
Proc. of the Winnipeg Wrokshop (2--6 August, 1994), S.~A. Fulling (ed),
Texas A\&M University, 1995,  423, (Hep-th/9606042).
\bibitem{diak}
D.V. Gal'tsov and O.V. Kechkin,
Phys. Lett. {\bf B361} 52 (1995), Phys. Rev.  {\bf D 54}, 1656 (1996).
\bibitem{er}
F.J. Ernst, Phys. Rev. {\bf 167}, 1175 (1968); {\bf 168}, 1415 (1968);
W. Kinnersley,  Journ. Math. Phys. {\bf 14}, 651 (1973);
{\bf 18}, 1529 (1977).
\bibitem{mg}
P.O. Mazur, Acta Phys. Polon. {\bf B14}, 219 (1983);
A. Eris, M. G\"urses, and A. Karasu, Journ. Math. Phys.
{\bf 25}, 1489 (1984).
\bibitem{bak}
I. Bakas,  Nucl. Phys. {\bf B428}, 374 (1994);
Phys. Lett. {\bf B343}, 103  (1995);
{\it Solitons in Axion--Dilaton Gravity}, CERN--TH--96--121,
(Hep-th/9605043).
\bibitem{bdg}
D.V. Gal'tsov, {\it Geroch-Kinnersley-Chitre group for 
Dilaton-Axion Gravity}, in ``Quantum Field Theory under the Influence of 
  External Conditions'', M. Bordag (Ed.) (Proc. of the International Workshop, 
  Leipzig, Germany, 18--22 September 1995), B.G. Teubner Verlagsgessellschaft, 
  Stuttgart--Leipzig, 1996, pp. 228-237, (Hep-th/9606041)

\bibitem{ad}
D.V. Gal'tsov, {\it Square of General Relativity},
A talk at the First Australasian Conference
on General Relativity and Gravitation, Adelaide,
February 12--17, 1996, to be published in the Proceedings;
DTP--MSU 96/14, Hep-th/9608021.
\bibitem{sd}
S. Shapere, S. Trivedi and F. Wilczek, Mod. Phys. Lett.
{\bf A6}, 2677, (1991);
A. Sen, Nucl. Phys. {\bf B404}, 109 (1993) ; Phys. Lett. {\bf 303B},
22 (1993) ; Int. J. Mod. Phys. {\bf A8}, 5079 (1993) ;
Mod. Phys. Lett. {\bf A8}, 2023 (1993) ;
J.H. Schwarz and A. Sen, Nucl. Phys. {\bf B411}, 35 (1994) ; Phys.
Lett. {\bf 312B}, 105 (1993).

\bibitem{iw} W. Israel and G.A. Wilson, Journal Math. Phys. {\bf 13}, 865
(1972).


\bibitem{nk}
G. Neugebauer and D. Kramer, Ann. der Physik (Leipzig)
{\bf 24}, 62 (1969).

\bibitem{ggk} D. Gal'tsov, A. Garc\'\i a and O. Kechkin, 
Journ. Math. Phys. {\bf 36}, 5023 (1995).
\bibitem{eh}
J. Ehlers, in {\em Les Theories Relativistes de la
Gravitation}, CNRS, Paris, 1959, p. 275.

\bibitem{gl}
D.V. Gal'tsov and P.S. Letelier,
{\it Interpolating black holes in dilaton--axion gravity},
Class. Quantum Grav. to appear (Gr-qc/9608023).

\bibitem{bgm}
P. Breitenlohner, D. Maison, and G. Gibbons,
Comm. Math. Phys. {\bf 120}, 253 (1988).

\bibitem{gkk}
R. Kallosh, M. Shmakova, and W. K. Wong, {\it``Freezing of Moduli by
N=2 Dyons,''}  SU--ITP--96--35, Hep-th/9607077;
G. Gibbons, R. Kallosh, and B. Kol, {\it Moduli, Scalar Charges, and the
First Law of Black Hole Thermodynamics}, SU--ITP--96--35, Hep-th/9607108.
\end{thebibliography}
\end{document}